\documentclass[11pt]{article}
\usepackage{graphicx}
\newcommand{\be}{\begin{equation}}
\newcommand{\ee}{\end{equation}}
\newcommand{\ba}{\begin{eqnarray}}
\newcommand{\ea}{\end{eqnarray}}

\newcommand{\n}{\label}

\begin{document}

\title{Symmetries and assisted inflation in the DGP braneworld model}

\author{Ruth Lazkoz \\
{\it \small Fisika Teorikoa eta Zientziaren Historia Saila,}\\
{\it \small Zientzia eta Teknologiaren Fakultatea, 
}\\
{\it \small  Euskal Herriko Unibertsitatea, 644 Posta Kutxatila, 48080 Bilbao, Spain}}
\date{\today}
\maketitle
\begin{abstract}

We present here the general transformation that leaves unchanged the form of
the field equations for perfect fluid cosmologies in the DGP braneworld model. 
Specifically, a prescription for relating exact solutions with different equations of state is provided, and the symmetries found can be used as algorithms for
generating new cosmological models from previously known ones. We also present, implicitly, the first known exact DGP perfect fluid spacetime.  A particular case of
the general transformation is used to illustrate  the crucial role played both by the number of scalar fields and the extra dimensional effects in the
occurrence of inflation. In particular, we see that assisted inflation does not proceed
at all times for one of the two possible ways in which the brane can be embedded into the bulk.

\end{abstract}

\maketitle
\section{Introduction}
Over the last years the brane world scenario has aroused  an overwhelming
curiosity (see \cite{review} and references therein). According to that idea our 4D universe lives on a domain wall embedded
in a manifold with  an  extra spatial dimension, so that gravity, but not standard models fields, can spread into this additional dimension.  In what cosmological implications are concerned, most attention has been devoted to the Randall Sundrum  (RS) II scenario \cite{RS}, which has recently faced the rise  of a challenger for the  title of the 
cosmologists' preferred brane cosmology model: the alternative scenario
proposed by Dvali, Gabadadze and Porrati (DGP) \cite{DGP}. In those two models the extra 
dimension is infinitely large. The  DGP proposal rests on the key assumption of the presence a four-dimensional Ricci scalar in the action. There are two main reasons that make this model phenomenologically appealing. First, it predicts that 4D Newtonian gravity on a brane world is regained at distances shorter than a given crossover scale $r_c$ (high energy  limit), whereas 5D effects become manifest above that scale~\footnote{  If the 
successful predictions of ordinary cosmology are to be preserved, the crossover scale $r_c$   must be of the order of  the Hubble scale today.} \,(low energy limit). Second, the model can explain late-time acceleration without having to invoke a cosmological constant or quintessential matter. Interestingly enough, the only way to mimic such scenario in a purely 4D is to introduce in the Einstein-Hilbert action covariant
terms with fractional powers of the Ricci scalar \cite{DDG}. Since that seems unjustifiable at present, cosmic
acceleration in the model is regarded as an intrinsically high-dimensional effect.

Geometrical aspects of the model have been studied in \cite{DDG,geometry}, and we will briefly review them in Section 2. In connection
with this, the remark must be done two different solution sets arise  depending 
on how the brane is embedded into the bulk, the discrepancy the sign in one of the terms
in the modified Friedmann equation.

Other studies
have focused on cosmological consequences \cite{crossover,Deffayet,cosmology}. In this work mathematical and physical aspects will overlap: First, in Section 3 we will formally show  how,
by exploiting symmetry properties, one can relate different  exact solutions to the evolution and conservation equations  equations describing perfect fluid FRW spacetimes in the DGP model. The procedure will establish
the link between cosmologies with different equations of state, and can be thought
of as an unusual solution generating algorithm. It must be stressed that, by construction, the new scale factor, Hubble factor, energy density and pressure will satisfy the same   
evolution and conservation equations as the seed model, and so such set of scalar quantities represent  our transformed solution. 

Next, in Section 5 we will apply our method to the generation of
inflationary solutions from non-inflationary ones. Specifically, the energy density of the transformed model will be a number of times larger than the original one, in the fashion 
of assisted inflation \cite{Liddle}. When applied to inflationary solutions, provided the energy density of the transformed model 
is large enough, those transformations will
give solutions with faster expansion or slower expansion, depending on which branch of solutions is being considered. Obviously, only in the case in which expansion is enhanced it will be possible to say that we have a realization of the assisted inflation phenomenon.
Then, a somehow unwanted  effect of the fifth dimension is that an increase in the energy density does not generically result in more acceleration.

Our discussion of assisted inflation is grounded on the same procedure as  in earlier papers \cite{precursor}, but the  reader may also be interested in looking at the different views on  brane assisted inflation taken in 
\cite{brane_assisted}.
Interestingly enough, assisted inflation produced by tachyonic instead of standard scalar fields
has also been considered \cite{tachyon}.

The possibilities of the method will be illustrated with exact examples. In particular,  we will consider exact perfect fluid FRW  cosmologies. Remarkably, exact DGP cosmologies have not appeared  in the literature before. Note, anyway, that the fact the solutions we present can only be given  implicitly does not make them less illustrative. For completeness, we will discuss the asymptotic behaviour of the solutions, and establish 
comparisons with related works. After that, we will apply the aforementioned transformations which relate inflationary and non-inflationary solutions, and will explicitly discuss in detail under which circumstances  assisted inflation occurs.

Finally, we will outline our main conclusions in Section 6.

\section{Basic  equations and main features} 
We now concentrate on the dynamics of a brane-world 
filled a perfect fluid with 
energy density $\rho$ and identical pressure
 $p$ along the three spatial directions (isotropic
perfect fluid). The model is governed by the Friedmann equation \cite{Deffayet}
\begin{equation}\label{Frie}
	\sqrt{H^2+\frac{k}{a^2}}= \sqrt{\frac{\kappa^2\rho}{3}+\frac{1}{4r_c^2}}\pm\frac{1}{2r_c}\,,
\end{equation} 
and the energy conservation equation 
\begin{equation}
\n{cons}
\dot\rho+3H(\rho+p)=0\,,
\end{equation}
where, $H\equiv\dot a/a$ is the Hubble factor, as usual.  Although the model can be generalized
to include a non trivial  brane tension and a 5D cosmological constant, we will stick with our simplified picture. Since The second term in the r.h.s. of (\ref{Frie}) can either be positive or negative, depending on how the brane is embedded into the bulk, we will  be dealing with two  branches of solutions \cite{Deffayet}. We will refer to them as to
the $(+)$ and the $(-)$ branch respectively, and wherever some term in a equation can take
different signs it must be understood that the upper one corresponds to the $(+)$ branch.

The modified Friedmann equation is not as easy to handle as its relativistic version, it is convenient to define an effective energy density $\rho_{\rm eff}(\rho)$ so that
\begin{equation}
H^2+\frac{k}{a^2}=\kappa^2 \frac{\rho_{\rm eff}}{3}\label{eff}
\end{equation} 
{F}rom equations (\ref{cons})--(\ref{eff}) and
 we obtain the Raychaudhuri equation
\be
\n{Hp}
\dot H=-\frac{1}{2}\,\kappa^2\frac{\partial\rho_{\rm eff}}{\partial \rho}(\rho+p)+\frac{k}{a^2}
\ee

Obviously, the  cosmological dynamics is  bound to be influenced by quantum corrections, although,  
interestingly, the modifications
are not relevant in all  energy   regimes. Needless to say, since in most expanding models the energy density decreases in the course of expansion, typically,  high and low energy regimes correspond to
early and late times, respectively. At early times, it follows that
\begin{equation}
H^2+\frac{k^2}{a^2}=\frac{\kappa^2 \rho}{3}+{\cal O}(\sqrt{\rho}),
\end{equation} 
so we see situation is the same as in general relativity (GR). In contrast, 
DGP and GR late-time dynamics are substantially different. Indeed, for matter that redshifts faster than curvature it can be seen that
\begin{equation}
H=\frac{1\pm1}{2\,{\sqrt{r_c}}} + 
  \frac{k^2\,{\sqrt{r_c}}\,\rho }{3} + 
  {\cal{O}}(\rho^2) .
\end{equation}
In the 
$(+)$ branch,  any expanding model evolves into a de Sitter phase and
inflation is guaranteed at late times. This is a striking feature of the model because
it provides a possible answer for the currently observed acceleration, without having
to resort to dark energy. In the $(-)$ branch we find that
late time expansion is also faster than in the relativistic setup, and, therefore, the conditions for the occurrence of inflation are less restrictive. If we speak in terms
of perfect fluids, inflation depends on the value of the $p/\rho$ ratio of the model, but if we want to express the same idea using scalar fields and potentials, then we will say
that the occurrence of accelerated expansion depends on how steep the potential is. In general, one needs fairly flat potentials for inflation to proceed.

In the relativistic setup, the assisted inflation proposal \cite{Liddle} came to change those views: If the universe is filled  with a single field and its potential is very steep, inflation may not occur, but if there are many of those fields inflation the situation may get reversed, and inflation may occur, the only requirement being a large enough
number of such fields. In other words, the cooperation of scalar fields with potentials not flat enough to produce inflation on their own, may result in enough effective potential energy for accelerated expansion. The specific models used for illustrating that idea considered FRW universes filled with sets  of $n$ identical fields with identical self-interaction potentials, thus, the effective energy
density of the model would be $n$ times the energy density of a single field. Correspondingly, the Hubble factor would get multiplied by $\sqrt{n}$ \cite{Chimento}.
Almost a decade ago, Barrow and Parsons \cite{Barrow} noticed the existence of this particular form-invariance transformation 
of the Einstein-Klein-Gordon equation system, although there the result  was neither interpreted  in terms form-invariance nor of assisted inflation.

This suggests that  assisted inflation may be equally successful beyond general relativity. 
For instance, that has been proved to be the case in the Randall-Sundrum model. Now, we give one further step and show in the next section that it also may also apply to the DGP model, provided certain conditions hold.
 
\section{Form-invariance transformations}
Our main purpose is to show how the scale factor changes when the energy density of  matter gets multiplied by a numerical factor. 
Given that the energy density  is a function of the scale factor, the whole transformation comes down to a suitable change in the scale factor. Note that the transformation must meet the consistency criterion that the transformed quantities correspond to another exact solution of the gravitational field equations. The discussion, however,  will be not restricted to any particular transformation. We will rather find out how the energy density, pressure, Hubble factor and deceleration factor get modified under an arbitrary change in the scale factor
subject to the requirement that the transformation leaves invariant the field equations \footnote{We insist on the fact that this is another way of saying that the transformed cosmological model will be a solution of those   equations}.

 We begin by considering a perfect fluid with energy density $\rho$ and pressure $p$, which is governed by equations (\ref{Frie})--(\ref{cons}). Given
a different perfect fluid with energy density $\bar\rho$ and pressure $\bar p$,
the corresponding equations will take the form
\begin{equation}
\label{00b}
\sqrt{\bar H^2+\frac{k}{\bar a^2}}= \sqrt{\frac{\kappa^2\bar \rho}{3}+\frac{1}{4r_c^2}}\pm\frac{1}{2r_c}\,,
\end{equation} 
\be
\n{cob}
\dot{\bar\rho}+3\bar H(\bar\rho+\bar p)=0.
\ee

As said above, the objective is to obtain a transformation that leaves the form of the 
system of equations (\ref{00b})--(\ref{cob}) unchanged. Put another way, 
we want to find a symmetry transformation that maps
(\ref{00b})--(\ref{cob}) into (\ref{Frie})--(\ref{cons}). Since we 
mean to obtain this transformation explicitly,  we make the Ansatz 
{\setlength\arraycolsep{0.1pt}
\ba
\n{ta}
\bar a \,&=&\bar a(a,H,\rho,p),\\
\n{th}
\bar H &=&\bar H\!(a,H,\rho,p),\\
\n{tr}
\bar\rho \,&=&\bar\rho(a,H,\rho,p),\\
\n{tp}
\bar p \,&=&\bar p(a,H,\rho,p).
\ea }

The  transformed versions of the last two equations can be obtained if we introduce an effective transformed energy density $\bar\rho_{\rm eff}(\bar\rho)$. We then get
\begin{equation}
\bar H^2+\frac{k}{\bar a^2}=\kappa^2 \frac{\bar\rho_{\rm eff}}{3}
\end{equation} 
and
\be
\n{Hpb}
\dot{\!\bar H}=-\frac{1}{2}\,\kappa^2\frac{\partial\bar\rho_{\rm eff}}{\partial  \bar \rho}(\bar\rho+\bar p)+\frac{k}{\bar a^2}.
\ee

In the procedure we are going to follow, we will regard as  symmetry transformations those which do not impose restrictions on
the functions appearing in the equations. When differentiating Eq. (\ref{th}), a term
proportional to $\dot p$ arises, but such term must vanish
identically, so that when  we insert that expression in the
l.h.s.~of Eq. (\ref{Hpb}) we get a new expression consistent with  the absence of terms proportional to $\dot p$ in the  r.h.s. of the same equation.
Then, necessarily $\bar H=\bar H(a,H,\rho)$. If we
replace now Eqs. (\ref{ta}) and (\ref{tr}) in Eqs. (\ref{00b}) and (\ref{cob}), and do
the same reasoning, we draw the conclusion that $\bar a=\bar a(a,H,\rho)$ and
$\bar\rho=\bar\rho(a,H,\rho)$. 

The next step is to calculate $\bar H$  from Eq. (\ref{ta})  using the
definition $\bar H=\dot {\bar a}/\bar a$, so that
\be
\n{bh}
\bar H=\frac{a}{\bar a}\frac{\partial\bar a}{\partial a}H+\frac{1}{\bar a}
\frac{\partial\bar a}{\partial H}
\left[\frac{k}{a^2}-\frac{1}{2}\kappa^2\frac{\partial\rho_{\rm eff}}{\partial\rho}(\rho+p)\right]-\frac{3H(\rho+p)}{\bar a}
\frac{\partial\bar a}{\partial \rho},
\ee
where equations (\ref{cons}) and (\ref{Hp}) have been used. Since $\bar H$
does not depend on $p$,  the coefficient of $p$ in Eq. (\ref{bh}) must vanish
as well. For that reason,
\be
\n{cp}
\kappa^2\frac{\partial\bar a}{\partial \left(3H^2\right)}+
\frac{\partial\bar a}{\partial(\rho_{\rm{eff}})}=0,
\ee
and the  general solution to the latter  is 
$\bar a=\bar a\left(a,\kappa^2\rho_{\rm{eff}}-3H^2\right)=
\bar a\left(a,3k/a^2\right)$, that
is,  $\bar a$ depends on $a$ only.

Summarizing, the transformation turns out
to be
\be
\n{taf}
\bar a=\bar a(a),
\ee
\be
\n{thf}
\bar H=\frac{\partial\ln \bar a}{\partial\ln a}\, H,
\ee
\be
\n{trf}
\bar\rho_{\rm{eff}}=\frac{3}{\kappa^2}\left[\left(\frac{\partial\ln \bar a}{\partial\ln a}\right)^2H^2+
\frac{k}{\bar a^2}\right],
\ee
\be
\n{tpf}
\bar p=-\bar\rho-\frac{2}{\kappa^2(\partial\rho_{\rm{eff}}/\partial\rho)}\left(\,\dot{\!\bar H}-\frac{k}{\bar a^2}\right),
\ee
where $\dot{\bar H}$ has to be calculated using Eq. (\ref{thf}).
Clearly, once the prescription $\bar a=\bar
a(a)$ is made the symmetry transformation gets completely determined.
Note that Eqs.(\ref{taf})-(\ref{tpf}) have been deduced without making any assumption on the equation of
state of the fluid.

Nevertheless, the $k=0$ case is more subtle, because since $a$ will neither appear explicitly in 
(\ref{Frie}) nor in (\ref{Hp}), the prescription to be made  is actually $\bar \rho(\rho)$,
and the pair of equations (\ref{thf}) and (\ref{trf}) will have to be replaced just by the single equation
\be
\n{thfb}
\bar H=\left(\frac{\bar\rho_{\rm{eff}}}{\rho_{\rm{eff}}}\right)^{1/2}H,
\ee
\section{Inflation from form-invariance}
Stages of accelerated expansion are marked by negative values of the deceleration parameter
\begin{equation} \label{q}
q (t)=-\frac{\ddot{a}}{aH^2}.
\end{equation}
%
Alternatively, one can write
\be
q+1=\frac{3}{2}\frac{\partial \log\rho_{\rm eff}}{\partial \rho}(\rho+p),
\ee
which, upon specific knowledge of  $\rho_{\rm eff}$, allows to say whether for given
$\rho$ and $p$ inflation occurs. Safely, in the GR case the requirement reduces to the
violation of the strong energy condition $\rho+3p>0$.

The transformation rule for $q$ is to be deduced from (\ref{thf}) or (\ref{thfb}) depending on whether we are considering cases with curvature or not, and in general one will obtain a 
complicated expression. Nevertheless, if we restrict ourselves (as we will in what follows) to cases in which $\bar a$, then (\ref{thf}) will hold both in the curved and non-curved cases, and it can be seen that
\be
\n{tq}
\bar q+1=\left[\frac{\partial\ln \bar a}{\partial\ln a}\right]^{-1}(q+1)+
\frac{\partial}{\partial\ln a}
\left[\frac{\partial\ln \bar a}{\partial\ln a}\right]^{-1}.
\ee

Let us concentrate for a moment on the $k\ne 0$ case. Application of our transformation  procedure  which, in principle, be able to tell us which $\bar a(a)$ rule gives some some desired
behaviour in either $\bar H$, $\bar \rho$, $\bar p$ or a related quantity. In previous works, we were concerned with the case in which $\bar H$ was a number of times larger
than $H$, that is $\bar H=nH$, because that would mean that a non-inflationary solution would be transformed into an inflationary one for a conveniently chosen proportionality factor $n$. Clearly, that situation arises when $\bar a=a^n$.

Alternatively, in the $k=0$ case, one starts by saying which is the relation between $\bar \rho$ and $\rho$ so that we reproduce some behaviour of our choice. Note that it is not possible to make a priori assumptions on the link
between $a$ and $a$, one rather has to find it by integration. Provided $\rho$ is a known function of $a$ consistent with the energy conservation equation,
and once  the functional dependence between $\rho$ and $\rho$ has been specified,  expressions of $\rho_{\rm{eff}}$ and $H$  in terms of $a$ would have to inserted in (\ref{thf}) to yield
an ordinary differential equation with $\bar a$ and $a$ as dependent and independent variables respectively. Integration of the latter
 would finally provide the $\bar a(a)$ rule leading to the pursued behaviour. 
 
Roughly speaking,  that approach to the  problem we just discussed was, in fact,  the
way in which assisted inflation was analyzed originally. As said above, in \cite{Liddle}
it was studied how the  deceleration factor changes under the multiplication by a numerical factor of the energy density of the scalar fields driving inflation.

For the sake of simplicity we are concentrating on flat models ($k=0$).  It follows that, in order to have
 $\bar\rho=n^2\rho$,  one  needs
\begin{equation}
\ln \bar a={\int _{1}^{a}\frac{3 \pm{\sqrt{\displaystyle 9 + 12\,n^2{\kappa }^2{{r_c}}^2\,\rho (u)}}}
       {u\,\left( 3 \pm {\sqrt{\displaystyle 9 + 12\,{\kappa }^2{{r_c}}^2\rho (u)}} \right) }\,du} \label{assisted},
\end{equation}
where integration constants has been
fixed so that 
$
\lim_{\rho\to\infty}\bar a=a^n $ 
and $\exists \,\lim_{\rho\to0}\bar a$, and because of those choices 
 \begin{equation}\displaystyle
\lim_{\rho\to 0} \ln\bar a= \left\{\begin{array}{ll}
\displaystyle\ln a&\mbox{ for the $(+)$ branch}\\
{n^2}\ln a&\mbox{  for the $(-)$ branch}
\end{array} \right..
\end{equation}

If we now insert (\ref{assisted})  in (\ref{trf}) and (\ref{tpf}), we would obtain the transformation rule for the energy density and pressure.Since in the DGP model the Friedmann equation is not linear in $\rho$,even though we are a case in which $\bar\rho$ depends linearly on $\rho$ $\bar H$ will not have a linear dependence on $H$; except in the 
high energy regime ($\rho\to\infty$). 

In the next section , we will turn to explicit examples, and will discuss whether assisted inflation proceeds. Neverthelss,  at this stage,  we have enough information so as to
discuss in broad terms whether our transformation enhances expansion or not. Eq. (\ref{assisted}) was obtained
upon integration of 
\begin{equation} \frac{\bar H}{H}= {\frac{3 \pm{\sqrt{\displaystyle 9 + 12\,n^2{\kappa }^2{{r_c}}^2\,\rho (a)}}}
       { 3 \pm {\sqrt{\displaystyle 9 + 12\,{\kappa }^2{{r_c}}^2\rho (a)}}  }}, \end{equation}which in turn was obtained by setting $\bar \rho=n^2\rho$ in (\ref{thf}).
       Let us assume $n>1$. It can be seen that for   $n^2\ge{\bar H}/{H}\ge n$. In contrast, for the $(+)$ branch $n\ge{\bar H}/{H}\ge 1$. When the seed solution reaches
       the de Sitter phase ($H=1$,$\rho=0$), so does the trasfomed solution.
One can see that the transition from $\bar H/H$ being closer to  $n$ (relativistic regime) than to $1$ (de Sitter regime) occurs at  $\rho=  {6\,\left( 1 + n \right) \,\left( 1 + 2\,n \right) }/
  ({\kappa\,{\left( 1 + 3\,n \right) }\,{{r_c}}} )^2$, and we infer the result  that the larger $n$ the larger the value of $\rho$ at which the transition occurs. In general,
        we see that  the expansion rate gets increased with the transformation in both branches, but the effect in more
accentuated in the $(-)$ branch. 
\section{Explicit examples}
So far, the discussion in this section has been carried out from a very broad perspective. 
It is time now to come to examples, and we choose
the illustrative $p=(\gamma-1)\rho$ case. From the energy conservation equation it follows that
\begin{equation}
 \rho=\frac{3}{4\kappa^2r_c^2}\left(\frac{a_0}{a}\right)^{3\gamma},
\end{equation}
$a_0$ being an arbitrary integration constant.
In the $k=0$ case exact solutions to the Friedmann equation can be given in the form $t=t(a)$:
{\setlength\arraycolsep{0.2pt}\begin{eqnarray}t&=&  
\frac{2r}{3\gamma}\left[v^{{\gamma }/{2}}\,\left(  {\sqrt{1 + v^{\gamma }}\mp v^{{\gamma }/{2}}} \right)  + 
  \log (v^{{\gamma }/{2}} + {\sqrt{1 + v^{\gamma }}})\right]
\end{eqnarray}}
where we have introduced a new variable $v=(a/a_0)^3$ for economy in the expressions.
Remarkably,  exact solutions to equations (\ref{Frie}) and (\ref{cons})
have not appeared in the literature before.

Let us discuss now some aspect of the asymptotic kinematics of the solutions. The extra dimensional effects do not show up at early times (high energy regime),  and, regardless of the branch, we
have
\begin{equation}
a\approx a_0\left(\frac{3\gamma t}{4r_c}\right)^{\frac{2}{3\gamma}},
\end{equation}
together with
$
q=-1+3\gamma/2$, which is exactly the same  result and in the relativistic framework. 
In contrast, at late times extra dimensional effects show up and  give  different solutions arise. For the $(+)$ branch we get
\begin{equation}a\approx a_0\left(\frac{1}{4}\,{\rm exp}\left(\frac{3\gamma t}{r_c}\right)\right)^{\frac{1}{3\gamma}}
\end{equation}
and $
q=-1$. This is, of course, the  solution  found by Deffayet in \cite{Deffayet}, which is self-inflationary because accelerated expansion
proceeds with a null energy density.
For the $(-)$ branch we get
\begin{equation}a\approx a_0\left(\frac{3 \gamma t}{4 r_c}\right)^{\frac{1}{3\gamma}}
\end{equation} and $q\approx-1+3\gamma$, and the condition for inflation becomes $\gamma>1/3$, 
 which is less restrictive than the relativistic condition ($\gamma>2/3$).
 
Obviously, asymptotic results are very relevant, but one may also be interested in the behaviour of the deceleration factor $q$ 
at any other time. In principle, it should be evaluated using (\ref{q}), but because that
is not an invertible expression, we have to resort
to yet one more alternative definition of $q$, namely
\begin{equation}
q=\displaystyle\frac{ a t''}{t'}
\end{equation}
where $'$ denotes differentiation with respect to $a$. We then obtain   \begin{equation}
q=-1 +\frac{3\gamma}{2} {\left( 1 \mp \frac{v^{{\gamma }/{2}} }
       {{\sqrt{1 + v^{\gamma }}}} \right) }
   \label{q_v}
\end{equation}
Combining (\ref{tq}) and (\ref{q_v}) one can obtain $\bar q$, which is a complicated expression made of two terms. The first one (see (\ref{tq})) is proportional to
\be
\frac{H}{\bar H}=\left[\frac{\partial\ln \bar a}{\partial\ln a}\right]^{-1}=\frac{\displaystyle 1 \pm {\sqrt{1 + {v }^{-\gamma }}}}
  {\displaystyle 1 \pm{\sqrt{1 + n^2\,{v }^{-\gamma }}}},
\ee
which is smaller than $1$ provided $n>1$, whereas the  other term is exactly
\be \frac{\partial}{\partial\ln a}\left[\frac{\partial\ln \bar a}{\partial\ln a}\right]^{-1}=
-\frac{3\,\gamma \,\left( 1 \pm{\sqrt{1 +{n^2}{v^{-\gamma }}}} - 
      n^2\,\left( 1 + {\sqrt{1 \pm v^{-\gamma }}}  \right)  \right) }{2\,v^{\gamma }\,
    {\sqrt{1 + v^{-\gamma }}}\,{\sqrt{1 + {n^2}{v^{-\gamma }}}}\,
    {\left( 1\pm{\sqrt{1 + {n^2}{v^{-\gamma }}}}   \right) }^2}\label{bit}.
\ee
In the $(-)$ branch, and for $n>1$, (\ref{bit}) is strictly negative, and it follows, then, that
\begin{equation}
(\bar q+1)<\left[\frac{\partial\ln \bar a}{\partial\ln a}\right]^{-1}(q+1)<(q+1),
\end{equation}
and  the transformed model accelerates faster than the original one.
\begin{figure}[t]
\label{figure:q plot}\centering
\includegraphics[width=.60\textwidth]{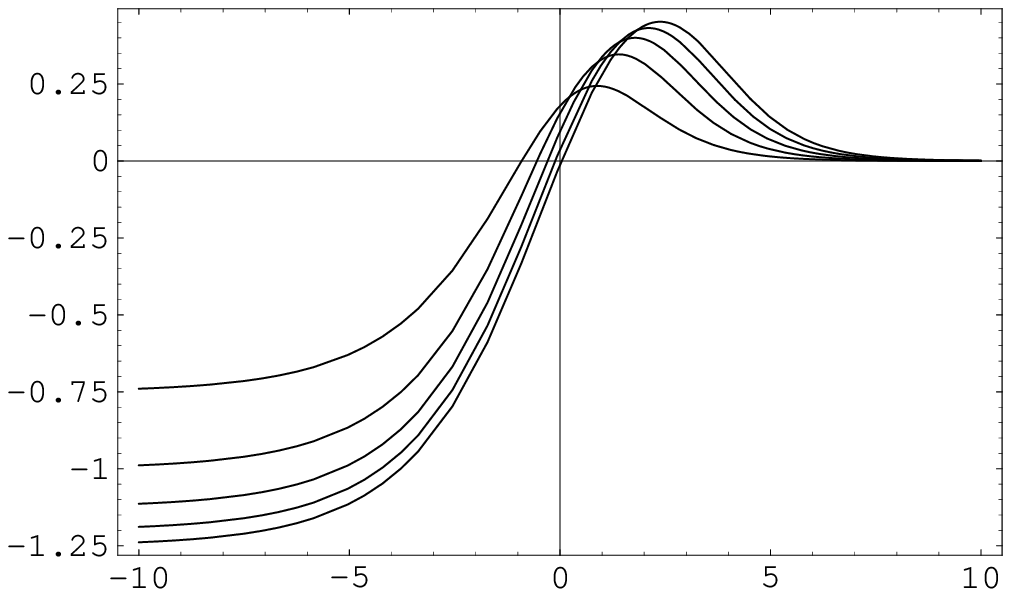}
\put(-108,-12){\small $\log v^{\gamma}$}
\put(-250,70){\small $\displaystyle\frac{\bar q-q}{\gamma}$}
\caption{Plot of ${(\bar q-q)}/{\gamma}$ as a function of $\log v^{\gamma}$ for $n=2,3,4,5,6$. The largest $n$, the largest the height of the peak.}
\end{figure}
However, the situation is not so clear cut for the $(+)$ branch, because in that case 
for $n>1$ (\ref{bit}) is  strictly positive. A series expansion in powers of $v^{\gamma}$ 
indicates that $\bar q<q$  for $v^{\gamma}\to 0$, but  $\bar q\le q$  may occur at later evolution stages. In Fig. 1 we have plotted  $(\bar q-q)/\gamma
$ 
as a function of $\log v^{\gamma}$ for  $n=2,3,\dots,6$. As can be seen, it agrees with our conclusion
that  $\bar q<q$ is the rule at  early times but  does not hold at late times. Consistently, the figure also reflects the fact that $\lim_{t\to 0} \bar q=\lim_{t\to 0}q$, because all models end up in a de Sitter phase. Note that  the peak in Fig. 1  which marks the transition between growing and decreasing $(\bar q-q)/\gamma$, moves to the right (to larger values of the energy density) as $n$ grows, exactly as happens to the value which marks the transition from $\bar H/H$ being closer to  $n$ (relativistic regime) than to $1$ (de Sitter regime).

\section{Conclusions}
In this paper we have discussed form-invariance transformations in the setup of DGP brane cosmology. Specifically, we have found the transformation rules for the scalar magnitudes
that characterize the dynamics of perfect fluid cosmologies so that a seed exact solution to  the corresponding field equations is mapped into a new exact solution. The transformation is rather general, in the sense that it may be applied to any FRW cosmology (with or without curvature),
regardless of the form of the Friedmann equation, which depends on the theoretical framework under consideration. As particular cases it includes, then, the transformations
obtained in \cite{Chimento,precursor}.

Such transformation can be used, for instance, towards generating new exact solutions.
In particular, one can construct inflationary cosmologies taking noninflationary seeds.
Depending on how the brane-world is embedded into the fifth dimension, two solutions sets 
 arise, the so called $(+)$ and $(-)$ branches. 

We turn to consider then models with a barotropic equation of state. We explicitly construct the corresponding exact solution and then stick with it in the subsequent discussion. In general, an increase in the energy density,  will give faster expansion in the $(-)$ branch, but in the other one, the contrary may happen. If the energy density is associated with a scalar field (the inflaton), and we think of an increase in it as being linked to a larger number of inflatons,  we can safely
speak of assisted inflation in the $(-)$ branch, very much like in the relativistic setup. 
However, this is not the case in the $(+)$ branch, and we see then that extra-dimensional effect
make assisted inflation non-generic. In particular, in the $(+)$ branch assisted inflation
only works at early  times.

Summarizing, we have shown once again how important a role form-invariance transformations 
can play in the analysis and interpretation of exact solutions.

\section*{Acknowledgements}
Thanks to J.M. Aguirregabiria, J.A. Valiente Kroon,  and S. Jhingan  for comments and suggestions. This work has been carried out with the support of the Basque Government through fellowship BFI01.412, the Spanish Ministry of Science and Technology
jointly with FEDER funds through research grant  BFM2001-0988,
and the University of the Basque Country through research grant 
UPV00172.310-14456/2002.

\end{document}